%% file: main.tex
\documentclass[10pt,twocolumn,letterpaper]{article}

\usepackage[pagenumbers]{cvpr} 

\usepackage{pifont}
\usepackage{graphicx}
\usepackage{amsmath}
\usepackage{amssymb}
\usepackage{algorithm}

\usepackage{booktabs}
\usepackage{makecell}
\usepackage{mathtools}
\usepackage{algpseudocode}
\usepackage{pbox}
\usepackage{footmisc}
\usepackage{float}
\usepackage{multirow,multicol}
\usepackage{color,colortbl}
\usepackage[normalem]{ulem}
\input{commands.tex}
\definecolor{Gray}{gray}{0.9}
\newcolumntype{g}{>{\columncolor{Gray}}c}
\usepackage[pagebackref,breaklinks,colorlinks]{hyperref}

\usepackage[capitalize]{cleveref}
\crefname{section}{Sec.}{Secs.}
\Crefname{section}{Section}{Sections}
\Crefname{table}{Table}{Tables}
\crefname{table}{Tab.}{Tabs.}

\def\cvprPaperID{5205} 
\def\confName{CVPR}
\def\confYear{2023}

\begin{document}
\title{Structured Kernel Estimation for Photon-Limited Deconvolution}

\author{Yash Sanghvi, Zhiyuan Mao, Stanley H. Chan\\
School of Electrical and Computer Engineering, Purdue University\\
{\tt\small \{ysanghvi, mao114, stanchan\}@purdue.edu}
}
\maketitle

\begin{abstract}
Images taken in a low light condition with the presence of camera shake suffer from motion blur and photon shot noise. While state-of-the-art image restoration networks show promising results, they are largely limited to well-illuminated scenes and their performance drops significantly when photon shot noise is strong.

In this paper, we propose a new blur estimation technique customized for photon-limited conditions. The proposed method employs a gradient-based backpropagation method to estimate the blur kernel. By modeling the blur kernel using a low-dimensional representation with the key points on the motion trajectory, we significantly reduce the search space and improve the regularity of the kernel estimation problem. When plugged into an iterative framework, our novel low-dimensional representation provides improved kernel estimates and hence significantly better deconvolution performance when compared to end-to-end trained neural networks. The source code and pretrained models are available at \url{https://github.com/sanghviyashiitb/structured-kernel-cvpr23}
\end{abstract}

\section{Introduction}

\textbf{Photon-Limited Blind Deconvolution}: This paper studies the photon-limited blind deconvolution problem. Blind deconvolution refers to simultaneously recovering both the blur kernel and latent clean image from a blurred image and "photon-limited" refers to presence of photon-shot noise in images taken in low-illumination /  short exposure. The corresponding forward model is as follows:
\begin{align}
    \vy = \text{Poisson}(\alpha \vh \circledast \vx). 
\end{align}
In this equation, $\vy \in \R^N$ is the blurred-noisy image, $\vx \in \R^N$ is the latent clean image, and $\vh \in \R^M$ is the blur kernel. We assume that $\vx$ is normalized to $[0,1]$ and the entries of $\vh$ are non-negative and sum up to $1$. The constant $\alpha$ represents the average number of photons per pixel and is inversely proportional to the amount of Poisson noise. 

\begin{figure*}[ht]
    \centering
    \begin{tabular}{ccccc}         \makecell{\includegraphics[width=0.19\linewidth,angle=90,origin=c]{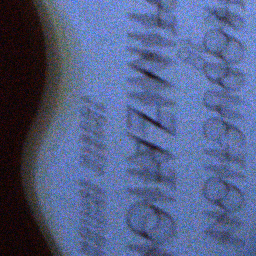} \\ \small Blurred and Noisy} &
    \hspace{-2.0ex}\makecell{\includegraphics[width=0.19\linewidth,angle=90,origin=c]{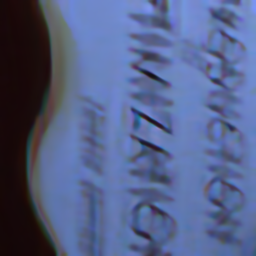} \\ MPR-Net \cite{zamir2021multi}  } &
    \hspace{-2.0ex}\makecell{\includegraphics[width=0.19\linewidth,angle=90,origin=c]{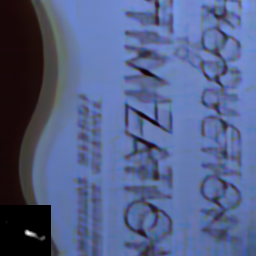} \\ \small Sanghvi et. al \cite{sanghvi2022photon}} &
    \hspace{-2.0ex}\makecell{\includegraphics[width=0.19\linewidth,angle=90,origin=c]{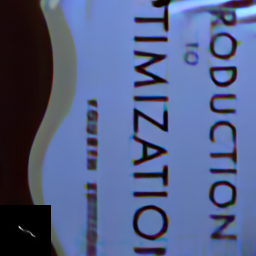} \\ \small \textbf{Ours} } & 
    \hspace{-2.0ex}\makecell{\includegraphics[width=0.19\linewidth,angle=90,origin=c]{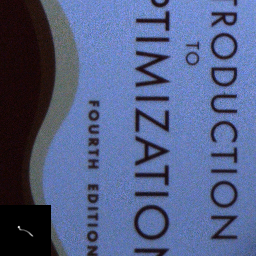} \\ \small Ground-Truth } 
    \end{tabular}
    \caption{\textbf{The proposed Kernel Trajectory Network (KTN) on real noisy blurred image from Photon-Limited Deblurring Dataset (PLDD) \cite{sanghvi2021photon}} The result corresponding to MPR-Net was generated by retraining the network with GoPro dataset \cite{Nah_2017_CVPR} corrupted by Poisson noise. The inset images for "Sanghvi et. al" and "Ours" represent the estimated kernel and the inset image for "Ground-Truth" represents the kernel captured using a point source, as provided in PLDD.}
    \label{fig:introduction}
\end{figure*}

\textbf{Deep Iterative Kernel Estimation}: Blind image deconvolution has been studied for decades with many successful algorithms including the latest deep neural networks \cite{Nah_2017_CVPR,zhang2019deep,zamir2021multi,tao2018scale,cho2021rethinking}. Arguably, the adaptation from the traditional Gaussian noise model to the photon-limited Poisson noise model can be done by retraining the existing networks with appropriate data. However, the restoration is not guaranteed to perform well because the end-to-end networks seldom explicitly take the forward image formation model into account. 

Recently, people have started to recognize the importance of blur kernel estimation for photon-limited conditions. One of these works is by Sanghvi et. al \cite{sanghvi2022photon}, where they propose an iterative kernel estimation method to backpropagate the gradient of an unsupervised reblurring function, hence to update the \emph{blur kernel}. However, as we can see in Figure \ref{fig:introduction}, their performance is still limited when the photon shot noise is strong. 

\textbf{Structured Kernel Estimation}: Inspired by \cite{sanghvi2022photon}, we believe that the iterative kernel estimation process and the unsupervised reblurring loss are useful. However, instead of searching for the kernel directly (which can easily lead to local minima because the search space is too big), we propose to search in a low-dimensional space by imposing structure to the motion blur kernel. 

To construct such a low-dimensional space, we frame the blur kernel in terms of trajectory of the camera motion. Motion trajectory is often a continuous but irregular path in the two-dimensional plane. To specify the trajectory, we introduce the concept of \emph{key point estimation} where we identify a set of anchor points of the kernel. By interpolating the path along these anchor points, we can then reproduce the kernel. Since the number of anchor points is significantly lower than the number of pixels in a kernel, we can reduce the dimensionality of the kernel estimation problem. 

The key contribution of this paper is as follows: We propose a new kernel estimation method called \emph{Kernel Trajectory Network (KTN)}. KTN models the blur kernel in a low-dimensional and differentiable space by specifying key points of the motion trajectory. Plugging this low-dimensional representation in an iterative framework improves the regularity of the kernel estimation problem. This leads to substantially better blur kernel estimates in photon-limited regimes where existing methods fail.

\section{Related Work}

\textbf{Traditional Blind Deconvolution}: Classical approaches to the (noiseless) blind deconvolution problem \cite{chan1998total,shan2008high,cho2009fast,xu2010two,mao2020} use a joint optimization framework in which both the kernel and image are updated in an alternating fashion in order to minimize a cost function with kernel and image priors. For high noise regimes, a combination of $\ell_1$+TV prior has been used in \cite{anger2019efficient}. Levin et. al \cite{levin2006blind} pointed out that this joint optimization framework for the blind deconvolution problem favours the no-blur degenerate solution i.e. $(\vx^*, \vh^*) = (\vy, \mI) $ where $\mI$ is the identity operator.  Some methods model the blur kernel in terms of the camera trajectory and then recover both the trajectory and the clean image using optimization 
\cite{whyte2012non,whyte2014deblurring,gupta2010single} and supervised-learning techniques \cite{zhang2021exposure,sun2015learning,gong2017motion}. 

For the non-blind case, i.e., when the blur kernel is assumed to be known, the Poisson deconvolution problem has been studied for decades starting from Richardson-Lucy algorithm \cite{richardson1972bayesian,lucy1974iterative}. More contemporary methods include Plug-and-Play \cite{rond2016poisson,sanghvi2021photon}, PURE-LET \cite{purelet}, and MAP-based optimization methods \cite{figueiredo2009deconvolution,harmany2011spiral}.

\textbf{Deep Learning Methods.} Recent years, many deep learning-based methods \cite{chakrabarti2016neural,schuler2015learning} have been proposed for the blind image deblurring task. The most common strategy is to train a network end-to-end on large-scale datasets, such as the GoPro \cite{Nah_2017_CVPR} and the RealBlur \cite{realblur} datasets. Notably, many recent works \cite{zamir2021multi,tao2018scale,Nah_2017_CVPR,cho2021rethinking,zhang2019deep} improve the performance of deblurring networks by adopting the multi-scale strategies, where the training follows a coarse-to-fine setting that resembles the iterative approach. Generative Adversarial Network (GAN) based deblurring methods \cite{kupyn2018deblurgan,Kupyn_2019_ICCV,Asim_2020_TIP,zhang2020deblurring} are also shown to produce visually appealing images. Zamir et al. \cite{Zamir2022Restormer} and Wang et al. \cite{Wang_2022_CVPR} adapt the popular vision transformers to the image restoration problems and demonstrate competitive performance on the deblurring task. 

\textbf{Neural Networks and Iterative Methods}: While neural networks have shown state-of-the-art performance on the deblurring task, another class of methods incorporating iterative methods with deep learning have shown promising results. Algorithm unrolling \cite{monga2021algorithm}, where an iterative method is unrolled for fixed iterations and trained end-to-end has been applied to image deblurring \cite{li2019algorithm,Agarwal2020}. In SelfDeblur \cite{Ren2020}, authors use Deep-Image-Prior \cite{ulyanov2018deep}  to represent the image and blur kernel and obtain state-of-the-art blind deconvolution performance.

\section{Method}
\subsection{Kernel as Structured Motion Estimation}
\begin{figure}
    \centering
    \includegraphics[page=1,trim = 20 5 25 10, width=0.90\linewidth]{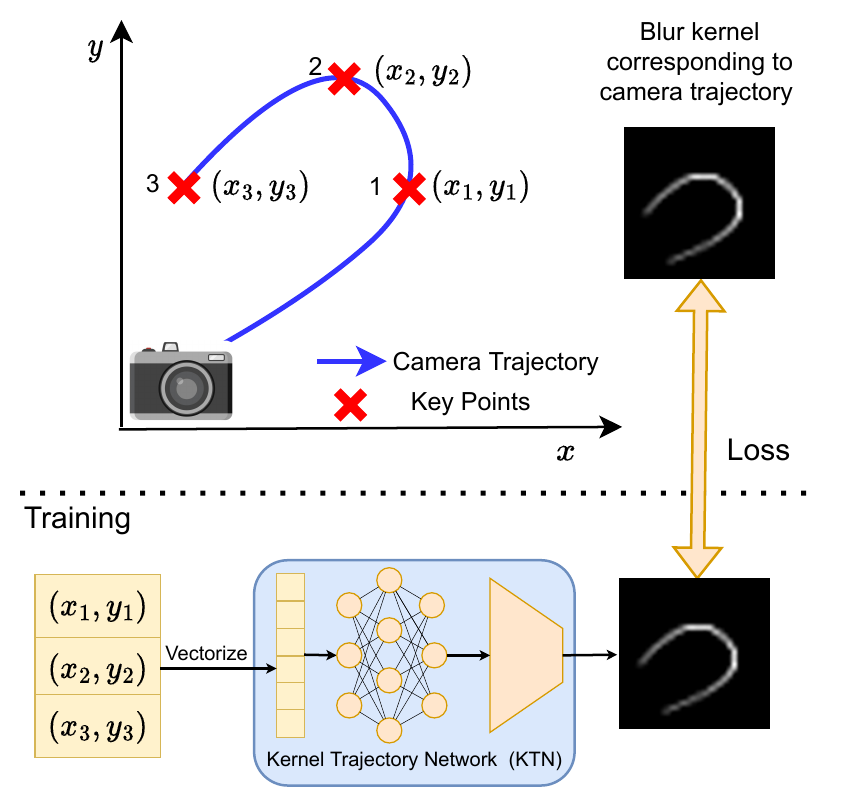}
    \caption{\textbf{Blur as Structured Motion Estimation}: In our formulation, we view the blur kernel as the continuous camera trajectory reduced to  $K$ key points, as shown in top half of the figure. We learn a differentiable representation from the vectorized $K$ key points to a blur kernel using a neural network. This \textit{lower dimensional and differentiable representation} is leveraged to estimate a better blur kernel and avoiding local minima during inference.}
    \label{fig:kernel_to_network}
\end{figure}
Camera motion blur can be modeled as a latent clean image $\vx$ convolved with a blur kernel $\vh$. If we assume the blur kernel lies in a window of size $32 \times 32$, then $\vh \in \R^{1024}$. However, in this high dimensional space, only few entries of the blur kernel $\vh$ are non-zero. Additionally, the kernel is generated from a two-dimensional trajectory which suggests that a simple sparsity prior is not sufficient. Given the difficulty of the photon-limited deconvolution problem, we need to impose a stronger prior on the kernel. To this end, we propose a \emph{differentiable and low-dimensional representation} of the blur kernel, which we will use as the search space in our kernel estimation algorithm. 

We take the two-dimensional trajectory of the camera during the exposure time and divide it into $K$ "key points". Each key point represents either the start, the end or a change in direction of the camera trajectory as seen in Figure \ref{fig:kernel_to_network}.  Given the $K$ key points as points mapped out in $x$-$y$ space, we can interpolate them using cubic splines to form a continuous trajectory in 2D. To convert this continuous trajectory to an equivalent blur kernel, we assume a point source image and move it through the given trajectory. The resulting frames are then averaged to give the corresponding blur kernel as shown in Figure \ref{fig:kernel_to_network}. 

Given the formulation of blur kernel $\vh$ in terms of $K$ key points, we now need to put this representation to a differentiable form since we intend to use it in an iterative scheme. To achieve this, we learn the transformations from the key points to the blur kernels using a neural network, which will be referred to as \emph{Kernel-Trajectory Network (KTN)}, and represent it using a differentiable function $T(.)$. Why differentiability is important to us will become clear to the reader in the next subsection. 

To train the Kernel-Trajectory Network, we generate training data as follows. First, for a fixed $K$, we get $K$ key points by starting from $(0,0)$ and choosing the next $K-1$ points by successively adding a random vector with a uniformly chosen random direction i.e. $U[0, 360]$ and uniformly chosen length from from $U[0, 100/(K-1)]$. Next, the set of key points are converted to a continuous smooth trajectory using bicubic interpolation. Then, we move a point source image through the given trajectory using the \verb|warpPerspective| function in OpenCV, and average the resulting frames.

Using the process defined above, we generate 60,000 blur kernels and their corresponding key point representations. For the Kernel-Trajectory Network $T(.)$,  we take a U-Net like network with the first half replaced by 3-fully connected layers and train it with the generated data using $\ell_2$-loss. For further architectural details on the KTN, we refer the reader to the supplementary document.

\subsection{Proposed Iterative Scheme}
\begin{figure*} 
\includegraphics[trim={0 0 20 0},clip,width=0.90\linewidth]{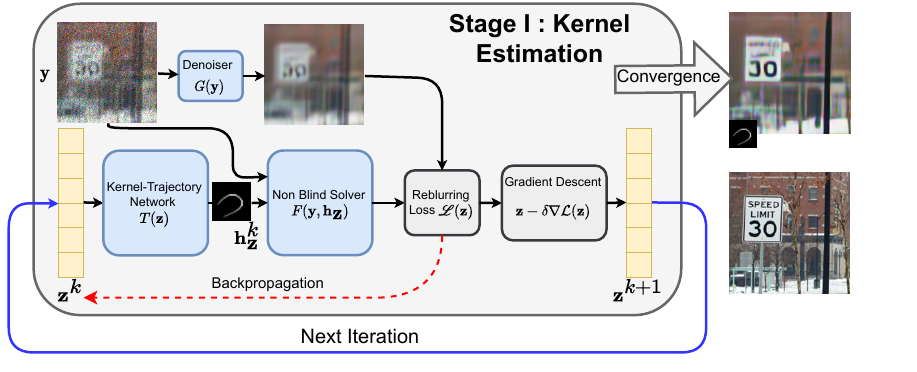}
    \caption{Flowchart describing first stage of the proposed scheme. We estimate the motion kernel of the blurry noisy image in lower dimensional latent space $\vz$ where the blur kernel is represented by $T(\vz)$ and by minimizing the reblurring loss $\mathcal{L}$ as defined in equation \ref{eq:optimization_stage_one}}
\end{figure*}
We described in the previous subsection how to obtain a low-dimensional and differentiable representation $T(.)$ for the blur kernel and now we are ready to present the full iterative scheme in detail. The proposed iterative scheme can be divided into three stages which are summarized as follows. We first generate an initial estimate of the direction and magnitude of the blur. This is used as initialization for a gradient-based scheme in Stage I which searches the appropriate kernel representation in the latent space $\vz$. This is followed by Stage II where we fine-tune the kernel obtained from Stage I using a similar process.

\textbf{Initialization}
Before starting the iterative scheme, we need a light-weight initialization method. This is important because of multiple local minima in the kernel estimation process.  

We choose to initialize the method with a rectilinear motion kernel, parameterized by length $\rho$ and orientation $\theta$. To determine the length and orientation of the kernel, we use a minor variation of the kernel estimation in PolyBlur \cite{delbracio2021polyblur}.  In this variation, the "blur-only image" $G(\vy)$ is used as the input and $\rho$, $\theta$ for the initial kernel  are estimated using the minimum of the directional gradients. We refer the reader to Section II in the supplementary document for further details on the initialization. Explanation on the "blur-only image" is provided when we describe Stage I of the scheme.

\textbf{Stage I: Kernel Estimation in Latent Space}
Given an initial kernel, we choose initial latent $\vz^0$ by dividing the rectilinear kernel into $K$ key points. 
Following the framework in \cite{sanghvi2022photon}, we run a gradient descent based scheme which optimizes the following cost function: 
\begin{equation}
    \calL(\vz) \bydef \underset{\text{Reblurring Loss}}{\underbrace{\| G(\vy) - \vh_{\vz} \circledast F(\vy,\vh_{\vz})) \|_2^2}}, 
\label{eq:optimization_stage_one}
\end{equation}
where $\vh_{\vz} \bydef T(\vz)$ represents the kernel output from Kernel-Trajectory network $T(.)$ given the vectorized key points representation $\vz$. $F(.)$ represents the  Poisson non-blind deconvolution solver which takes both noisy-blurred image and a blur kernel as the input. $G(\vy)$ represents a denoiser which is trained to remove \emph{only the noise from noisy-blurred image}. The overall cost function represents reblurring loss i.e. how well the kernel estimate and corresponding image estimate $\vh_{\vz} \circledast F(\vy, \vh_{\vz})$ match the blur-only image $G(\vy)$. 

To minimize the cost function in \eqref{eq:optimization_stage_one}, we use a simple gradient descent based iterative update for $\vz$ as follows:
\begin{align}
    \vz^{k+1} = \vz^k - \delta \underset{\text{backpropagation}}{\underbrace{\nabla_{\vz} \mathcal{L}(\vz^{k})}}   
\end{align}
where $\delta > 0$ is the step size and $\nabla_{\vz}\mathcal{L}(\vz^{k})$ represents the gradient of the cost function $\mathcal{L}$ with respect to $\vz$ evaluated $\vz^k$. It should be noted that the cost function is evaluated using the non-blind solver $F(.)$ and Kernel-Trajectory Network $T(.)$ - two neural network computations. Therefore, we can compute the gradient $\nabla_{\vz}\mathcal{L}(\vz^{k})$ using auto-differentiation tools provided in \verb|PyTorch| by backpropagating the gradients through $F(.)$ and then $T(.)$ 

\textbf{Stage II: Kernel Fine-tuning}
In the second stage, using the kernel estimate of Stage I, we fine-tune the kernel by "opening up" the search space to the entire kernel vector instead of parametrizing by $T(.)$.
Specifically, we optimize the following loss function
\begin{align}
    \calL(\vh) \bydef \| G(\vy) - \vh \circledast F(\vy,\vh)) \|_2^2 + \gamma \|\vh\|_1. \label{eq:optimization_stage_two}
\end{align}
Note the presence of the second term which acts as an $\ell_1$-norm sparsity prior. Also the kernel vector $\vh$ is being optimized instead of the latent key point vector $\vz$. Using variable splitting as used in Half-Quadratic Splitting (HQS), we convert the optimization problem in \eqref{eq:optimization_stage_two} to as follows:
\begin{align}
    \mathcal{L}(\vh,\vv) = \| G(\vy) - \vh \circledast F(\vy,\vh)) \|_2^2 + \gamma \|\vh\|_1 + \frac{\mu}{2}\|\vh-\vv\|_2^2
\end{align}
for some hyperparameter $\mu > 0$. This leads us to the following iterative updates 
\begin{align}
    \vh^{k+1} = \vh^k - \delta \cdot \big\{ {\nabla_{\vh}\mathcal{L}(\vh^k)} + \mu(\vh^k - \vv^k)\big\}, \\
    \vv^{k+1} 
    = \max\left(\left|\vh^{k+1}\right|-\gamma/\mu,0\right)\cdot \text{sign}(\vh^{k+1}) \notag \\ \bydef \calS_{\gamma/\mu}(\vh^{k+1}).
\end{align}

\begin{algorithm}[ht]
\begin{algorithmic}[1]
\State \textbf{Input}: Noisy-blurry $\vy$, Photon-Level $\alpha$, denoiser $G(\cdot)$, non-blind solver $F(\cdot)$, Kernel-Trajectory-Network $T(\cdot)$. 
\State Initialize $\vz^0$ using method described in Algorithm 1 from supplementary
    \For{$k = 0, 1, 2, \cdot\cdot\cdot$} \% \small \emph{Stage I begins here}
    \State $\vh_{\vz}^k \leftarrow T(\vz^k)$
    \State $\mathcal{L}(\vz) \leftarrow \|G(\vy) - \vh_{\vz}^k \circledast F(\vy,\vh_{\vz}^k)\|_2^2$
    \State Calculate $\nabla_{\vz}\mathcal{L}(\vz^k)$ using automatic differentiation
    \State $\vz^{k+1} \leftarrow \vz^{k} - \delta  \nabla_{\vz}\mathcal{L}(\vz^k)$
    \EndFor

\State $\vh^0, \vv^0 \leftarrow T(\vz^{\infty})$, $\mu \leftarrow 2.0$, $\gamma \leftarrow 10^{-4}$
    \For{$k = 0, 1, 2, \cdot\cdot\cdot$} \% \small \emph{Stage II begins here}
    \State $\mathcal{L}(\vh) \leftarrow \|G(\vy) - \vh \circledast F(\vy,\vh)\|_2^2$
    \State Calculate $\nabla_{\vh}\mathcal{L}(\vh^k)$ using automatic differentiation
    \State $\vh^{k+1} \leftarrow \vh^{k} - \delta \big( \nabla_{\vh}\mathcal{L}(\vh^k) + \mu(\vh^k - \vv^k)  \big)$
    \State $\vv^{k+1} \leftarrow \mathcal{S}_{\gamma/\mu}(\vh^{k+1})$
    \State $\mu \leftarrow 1.01 \mu$
    \EndFor
\State return $\vh^{(\infty)}$ and $\vx^{(\infty)} = F(\vy,\vh^{(\infty)})$
\end{algorithmic}
\caption{Iterative Poisson Deconvolution Scheme}
\label{alg:poiss_deconv}
\end{algorithm}

\section{Experiments}
\subsection{Training}
While our overall method is not end-to-end trained, it contains pre-trained components, namely the non-blind solver $F(.)$ and denoiser $G(.)$. The architectures of $F(.)$ and $G(.)$ are inherited from PhD-Net \cite{sanghvi2021photon} which takes as input a noisy-blurred image and kernel. For the denoiser $G(.)$, we fix kernel input to identity operator since it is trained to remove only the noise from noisy-blurred image $\vy$. 

$F(.)$ and $G(.)$ are trained using synthetic data as follows. We take clean images from  Flickr2K dataset \cite{Flickr2K} and the blur kernels from the code in Boracchi and Foi \cite{motion_blur}. The blurred images are also corrupted using Poisson shot noise with photon levels $\alpha$ uniformly sampled from $[1,60]$. The non-blind solver $F(.)$ is trained using kernel and noisy blurred image as input, and clean image as the target. The denoiser $G(.)$ is trained with similar procedure but with blurred-noisy image as the only input blur-only images as the target. The training processes, along with other experiments described in this paper are implemented using \verb|PyTorch| on a NVIDIA Titan Xp GPU. 

For quantitative comparison of the method presented, we retrain the following state-of-the-art networks for Poisson Noise: \emph{Scale Recurrent Network (SRN)} \cite{tao2018scale}, \emph{Deep-Deblur} \cite{Nah_2017_CVPR}, \emph{DHMPN} \cite{zhang2019deep}, \emph{MPR-Net} \cite{zamir2021multi}, and \emph{MIMO-UNet+} \cite{cho2021rethinking}. 
We perform this retraining in the following two different ways. First, we use synthetic data training as described for $F(.)$ and $G(.)$. Second, for testing on realistic blur, we retrain the networks using the GoPro dataset \cite{Nah_2017_CVPR} as it is often used to train neural networks in contemporary deblurring literature. We add the Poisson noise with the same distribution as the synthetic datasets to the blurred images. While retraining the networks, we use the respective loss functions from the original papers for sake of a fair comparison. 

\subsection{Quantitative Comparison}

\begin{table*}[h]
    \setlength\doublerulesep{0.5pt}
    \centering
    \hspace{-1.0ex}
    \scalebox{0.90}{
    \begin{tabular}{p{3em}ccccccc|ggg}
        \toprule
        \multicolumn{2}{c}{ \makecell{Method  \\ Photon Level, Metric}} & \makecell{SRN \\ \cite{tao2018scale}} & \makecell{ DHMPN  \\ \cite{zhang2019deep}} & \makecell{ Deep-\\Deblur \cite{Nah_2017_CVPR} } & \makecell{ MIMO-\\UNet+ \cite{cho2021rethinking}} & \makecell{MPRNet \\ \cite{zamir2021multi}} &  Ours & \makecell{ P4IP \\ \cite{rond2016poisson}} & \makecell{PURE-LET \\ \cite{purelet}}  & \makecell{PhD-Net \\ \cite{sanghvi2021photon}} \Tstrut \\
         \midrule[0.3pt] 
        \multirow{4}{*}[0.3em]{$\alpha = 10$} & PSNR $\uparrow$ & 20.71 & 20.89 & 21.17 & 21.04 & 21.09 & \textbf{21.57} & 19.26 & 22.49 & 23.00 \\
        & SSIM $\uparrow$ & 0.386 & 0.391 & 0.401 & 0.356 & 0.393 & \textbf{0.471} & 0.348 & 0.485 & 0.500 \\
         & LPIPS-Alex $\downarrow$ & 0.681 & 0.702 & 0.656 & 0.733 & 0.678 & \textbf{0.560} & 0.733 & 0.588 & 0.544 \\
         & LPIPS-VGG $\downarrow$ & 0.646 & 0.652 & 0.627 & 0.683 & 0.641 & \textbf{0.587} & 0.674 & 0.607 & 0.567 \\
         \midrule[0.3pt]
         \multirow{4}{*}[0.3em]{$\alpha = 20$} & PSNR $\uparrow$ & 20.79  & 21.03  & 21.30 & 21.36 & 21.25 & \textbf{21.93} & 19.45 & 22.94 & 23.63\\
        & SSIM $\uparrow$ & 0.392 & 0.401 & 0.410 & 0.396 & 0.405 & \textbf{0.483} & 0.353 & 0.516 & 0.540 \\
         & LPIPS-Alex $\downarrow$ & 0.683 & 0.688 & 0.666 & 0.660 & 0.667 & \textbf{0.542} & 0.726 & 0.526 & 0.500 \\
         & LPIPS-VGG $\downarrow$ & 0.639 & 0.640 & 0.621 & 0.663 & 0.631 & \textbf{0.578} & 0.668 & 0.584 & 0.539 \\
         \midrule[0.3pt]
         \multirow{4}{*}[0.3em]{$\alpha = 40$} & PSNR $\uparrow$ & 20.89 & 21.15 & 21.43 & \textbf{21.63} & 21.41 & 21.62 & 20.18 & 23.48 & 24.38 \\
        & SSIM $\uparrow$ & 0.409 & 0.418 & 0.425 & 0.441 & 0.428 & \textbf{0.527} & 0.372 & 0.561 & 0.593 \\
         & LPIPS-Alex $\downarrow$ & 0.677 & 0.673 & 0.673 & 0.586 & 0.647 & \textbf{0.488} & 0.706 & 0.467 & 0.446 \\
         & LPIPS-VGG $\downarrow$ & 0.629 & 0.626 & 0.612 & 0.639 & 0.613 & \textbf{0.549} & 0.660 & 0.557 & 0.503 \\
         \midrule[0.3pt]
         \multicolumn{2}{c}{Blind?} & \cmark & \cmark &  \cmark &\cmark & \cmark & \cmark & \xmark & \xmark & \xmark\\
        \multicolumn{2}{c}{End-To-End Trained?} & \cmark & \cmark & \cmark & \cmark & \cmark & \xmark & \xmark & \xmark & \cmark \\
        \bottomrule[1pt]
        \bottomrule[1pt]
    \end{tabular}
    }
    \vspace{1ex}
    \caption{\textbf{Performance on BSD100 Dataset with Synthetic Blur}. $\uparrow$ represents metrics where higher means better and vice versa for $\downarrow$. LPIPS-Alex and LPIPS-VGG represent the perceptual measures from \cite{lpips}. The best performing blind deconvolution method for each metric and photon level is shown in \textbf{bold}.  The non-blind deconvolution methods are shown for reference in grey columns.  }
    \label{tab:bsd100}
\end{table*}
\begin{table*}[ht]
    \setlength\doublerulesep{0.5pt}
    \centering
    \hspace{-1.0ex}
    \scalebox{0.90}{
    \begin{tabular}{p{3em}ccccccc|ggg}
        \toprule
        \multicolumn{2}{c}{ \makecell{Method \\ Photon Level, Metric}} & \makecell{SRN \\ \cite{tao2018scale}} & \makecell{ DHMPN  \\ \cite{zhang2019deep}} & \makecell{ Deep-\\Deblur  \cite{Nah_2017_CVPR} } & \makecell{ MIMO-\\UNet+ \cite{cho2021rethinking}} & \makecell{MPRNet \\ \cite{zamir2021multi}} &  Ours & \makecell{ P4IP \\ \cite{rond2016poisson}} & \makecell{PURE-LET \\ \cite{purelet}}  & \makecell{PhD-Net \\ \cite{sanghvi2021photon}} \Tstrut \\
         \midrule[0.3pt] 
        \multirow{4}{*}[0.3em]{$\alpha = 10$} & PSNR $\uparrow$ & 20.26 & 20.50 & 20.93 & 21.25 & 21.04  & \textbf{22.01} & 19.92 & 21.63 & 22.41  \\
        & SSIM $\uparrow$ & 0.510 & 0.509 & 0.524 & 0.516 & 0.533  & \textbf{0.611} & 0.463 & 0.590 & 0.638 \\
         & LPIPS-Alex $\downarrow$ & 0.507 & 0.521 & 0.496 & 0.594 & 0.479 & \textbf{0.340} & 0.546 & 0.371 & 0.341 \\
         & LPIPS-VGG $\downarrow$ & 0.531 & 0.526 & 0.518 & 0.661  & 0.511 & \textbf{0.477} & 0.555 & 0.522 & 0.466   \\
         \midrule[0.3pt]
         \multirow{4}{*}[0.3em]{$\alpha = 20$} & PSNR $\uparrow$ &  20.49 & 20.39 & 21.11 & 21.64 & 21.33 & \textbf{22.72} & 19.53 & 21.79 & 22.78  \\
        & SSIM $\uparrow$ & 0.523 & 0.521 & 0.536 & 0.554  & 0.551 & \textbf{0.641} & 0.442 & 0.607 & 0.667 \\
         & LPIPS-Alex $\downarrow$ & 0.496 & 0.502 & 0.492 & 0.485 & 0.459 & \textbf{0.304} & 0.533 & 0.339 & 0.304 \\
         & LPIPS-VGG $\downarrow$ &  0.515 &  0.514  & 0.501  & 0.610 & 0.493 & \textbf{0.448} & 0.554 & 0.510 & 0.447 \\
         \midrule[0.3pt]
         \multirow{4}{*}[0.3em]{$\alpha = 40$} & PSNR $\uparrow$ & 20.59 & 20.50 &  21.20 & 21.88 & 21.54 & \textbf{22.32} & 17.32 & 21.78 & 22.96  \\
        & SSIM $\uparrow$ & 0.535 & 0.532 & 0.545 & 0.583  & 0.567  & \textbf{0.647} & 0.362 & 0.614 & 0.687   \\
         & LPIPS-Alex $\downarrow$ & 0.491 & 0.494 & 0.494 &  0.428 & 0.447  & \textbf{0.273} & 0.487 & 0.324 & 0.263  \\
         & LPIPS-VGG $\downarrow$ & 0.506 & 0.506 & 0.493 & 0.557 & 0.479 & \textbf{0.444} & 0.560 & 0.507 & 0.432  \\
        \bottomrule[1pt]
        \bottomrule[1pt]
    \end{tabular}
    }
    \vspace{1ex}
    \caption{\textbf{Performance on Levin dataset with realistic camera shake blur} \cite{levin2006blind}. The best performing blind deconvolution method for each metric and photon level is shown in \textbf{bold} and non-blind deconvolution methods are shown for reference in grey columns. }
    \label{tab:levin}
\end{table*}
We quantitatively evaluate the proposed method on three different datasets, and compare it with state-of-the-art deblurring methods. In addition to the end-to-end trained methods described previously, we also compare our approach to the following Poisson deblurring methods: Poisson-Plug-and-Play \cite{rond2016poisson}, and PURE-LET \cite{purelet}. Even though these methods assume the blur kernel to be known, we include them in the quantitative comparison since they are specifically designed for Poisson noise. For all of the methods described above, we compare the restored image's quality using PSNR, SSIM, and Learned Perceptual Image Patch Similarity (LPIPS-Alex, LPIPS-VGG) \cite{lpips}. We include the latter as another metric in our evaluation since failure of MSE/SSIM to assess image quality has been well documented in \cite{wang2009mean,lpips}

\textbf{BSD100}: First, we evaluate our method on synthetic blur as follows. We collect 100 random images from the BSD-500 dataset, blur them synthetically with motion kernels from the Levin dataset \cite{levin2009understanding} followed by adding Poisson noise at photon-levels $\alpha = 10, 20$, and $40$.  The results of the quantitative evaluation are provided in Table \ref{tab:bsd100}. Since the blur is synthetic, ground-truth kernel is known and hence, can be used to simultaneously evaluate Poisson non-blind deblurring methods i.e, Poisson Plug-and-Play, PURE-LET, and PhD-Net. The last method is the non-blind solver $F(.)$ and serves as an upper bound on performance.

\textbf{Levin Dataset}: Next, we evaluate our method on the Levin dataset \cite{levin2009understanding} which contains 32 real blurred images along with the ground truth kernels, as measured through a point source. We evaluate our method on this dataset with addition of Poisson noise at photon levels $\alpha = 10, 20$ and $40$ and the results are shown in Table \ref{tab:levin}. For a fair comparison, end-to-end trained methods are retrained using synthetically blurred data (as described in Section IV-A) for evaluation on BSD100 and Levin dataset.

\textbf{RealBlur-J}\cite{realblur}: To demonstrate that our method is able to handle realistic blur, we evaluate our performance on randomly selected 50 patches of size $256 \times 256$ from the Real-Blur-J \cite{realblur} dataset. Note that we reduce the size of the tested image because our method is based on a single-blur convolutional model. Such model may not be applicable for a large image with spatially varying blur and local motion of objects. However, for a smaller patch of a larger image, the single-blur-kernel model of deconvolution is a much more valid assumption. 

To ensure a fair comparison, we evaluate end-to-end networks by  retraining on both the synthetic and GoPro datset. As shown in Table \ref{tab:psnr_ssim_realblur}, we find that end-to-end networks perform consistently better on the RealBlur dataset when trained using the GoPro dataset instead of synthetic blur. This can be explained by the fact both GoPro and RealBlur have realistic blur which is not necessarily captured by a single blur convolutional model.  

\begin{table*}[ht]]
    \setlength\doublerulesep{0.5pt}
    \centering
    \scalebox{0.90}{
    \begin{tabular}{c|c|cc|cc|cc|cc|cc|c}
        \toprule
        \multicolumn{2}{c}{ \makecell{Method $\rightarrow$ \\ }} & \multicolumn{2}{c}{SRN  \cite{tao2018scale}} & \multicolumn{2}{c}{ DHMPN  \cite{zhang2019deep}} & \multicolumn{2}{c}{Deep-Deblur \cite{Nah_2017_CVPR}} & \multicolumn{2}{c}{MIMO-UNet+\cite{cho2021rethinking}} & \multicolumn{2}{c}{MPRNet  \cite{zamir2021multi}} & \textbf{Ours} \Tstrut \\
         \midrule[0.3pt] 
         \multicolumn{2}{c}{\makecell{Training $\rightarrow$ \\ Photon lvl, Metric}} & Synth. & GoPro & Synth. & GoPro & Synth. & GoPro & Synth. & GoPro & Synth. & GoPro & Synth. \Tstrut \\
         \midrule[0.3pt] 
        \multirow{4}{*}[0.3em]{$\alpha = 10$} & PSNR $\uparrow$ & 25.72 & 27.64 & 25.72 & 27.58 & 25.98 & 27.57 &  26.20 & 26.78 & 26.26 & \textbf{28.16} & \underline{26.61} \\
        & SSIM $\uparrow$ & 0.612 & 0.706 & 0.603 & 0.696  & 0.577 & 0.719 & 0.531 & 0.571 & 0.641 & 0.729 & \underline{\textbf{0.738}} \\
         & LPIPS-Alex $\downarrow$ & 0.438  & 0.310 & 0.454 & 0.329 & 0.441 & 0.297 & 0.484 & 0.396 & 0.401 & 0.288 & \underline{\textbf{0.277}} \\
         & LPIPS-VGG $\downarrow$ & 0.508  & 0.454 & 0.509  & 0.472  & 0.496 &  0.440 & 0.549 & 0.508 & 0.496 & 0.427 & \underline{\textbf{0.416}} \\
        \midrule[0.3pt]
        \multirow{4}{*}[0.3em]{$\alpha = 20$} & PSNR $\uparrow$ & 25.37 & 27.91 & 25.46 & 28.02 & 25.95 & 27.81 & 26.69 & 27.53 & 26.51 & \textbf{28.29} & \underline{27.23} \\
        & SSIM $\uparrow$ & 0.658 & 0.775 & 0.655 & 0.764  & 0.636 & 0.778 & 0.630 & 0.678 & 0.715 & \textbf{0.793} & \underline{\textbf{0.793}} \\
         & LPIPS-Alex $\downarrow$ & 0.426 & 0.275 & 0.429 & 0.288 & 0.427 & 0.265 & 0.401 & 0.313 & 0.360 & 0.256 & \underline{\textbf{0.241}} \\
         & LPIPS-VGG $\downarrow$ & 0.492 & 0.421 & 0.496 & 0.437 & 0.485 & 0.410 & 0.495 & 0.446 & 0.466 & 0.402 & \underline{\textbf{0.382}} \\
         \midrule[0.3pt]
        \multirow{4}{*}[0.3em]{$\alpha = 40$} & PSNR $\uparrow$ & 25.67 & 28.34 & 25.72 & 28.27 & 26.22 &  28.13 & \underline{27.24} & 28.14 & 26.85 & \textbf{28.72} & 27.11 \\
        & SSIM $\uparrow$ & 0.665 & 0.768 & 0.653 & 0.760 & 0.626 & 0.771  & 0.675 & 0.712 & 0.716 & \textbf{0.788} & \underline{0.782} \\
         & LPIPS-Alex $\downarrow$ & 0.415 & 0.268 & 0.418 & 0.268 & 0.418 & 0.258 & 0.347 & 0.267 & 0.343 & 0.245 & \underline{\textbf{0.221}} \\
         & LPIPS-VGG $\downarrow$ & 0.482 & 0.405 & 0.481 & 0.413 & 0.470 & 0.396 & 0.457
         & 0.404 & 0.444 & 0.386 & \underline{\textbf{0.360}} \\
        \bottomrule[1pt]
        \bottomrule[1pt]
    \end{tabular}
    }
    \vspace{1ex}
    \caption{\textbf{Performance on RealBlur-J Dataset with realistic blur} \cite{realblur}: \textbf{Bold} and \underline{underline} refer to overall best performing method and best synthetic performance method. It should be noted that methods that are not trained end-to-end are usually at disadvantage when comparing on metrics like PSNR. However, it can be seen that our reconstruction is generally preferred by other perceptual metrics.}
    \label{tab:psnr_ssim_realblur}
\end{table*}

\subsection{Qualitative Comparison}
\begin{figure*}[ht]
    \centering
    \begin{tabular}{cccc}
 \hspace{-2.0ex}\multirow{2}[2]{*}[17mm]{\makecell{\includegraphics[width=0.39\linewidth]{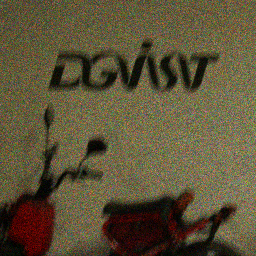} \\ Input }} &
\makecell{\includegraphics[width=0.18\linewidth]{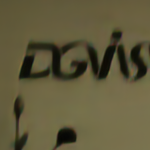} \\ SRN } &
\hspace{-1.5ex}\makecell{\includegraphics[width=0.18\linewidth]{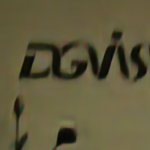} \\ DHMPN} & 
\hspace{-1.5ex}\makecell{\includegraphics[width=0.18\linewidth]{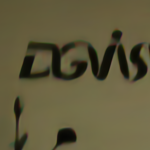} \\ MPR-Net} \\
 & \hspace{-1.5ex}\makecell{\includegraphics[width=0.18\linewidth]{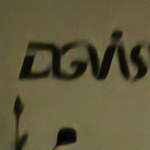} \\ MIMO-UNet+}  
 & \makecell{\includegraphics[width=0.18\linewidth]{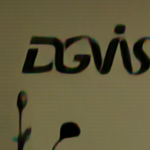} \\ \textbf{Ours}} & \hspace{-1.5ex}\makecell{\includegraphics[width=0.18\linewidth]{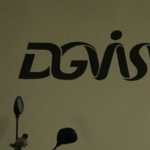} \\ Ground-Truth}   

\end{tabular}
\caption{\textbf{Qualitative example on the Real-Blur Dataset}: For a more extensive set of results, we refer the reader to the supplementary document.}
    \label{fig:real_blur_examples}
\end{figure*}

\begin{figure*}[ht]
\centering
\begin{tabular}{cccc}
 \hspace{-2.0ex}\multirow{2}[2]{*}[17mm]{\makecell{\includegraphics[width=0.39\linewidth]{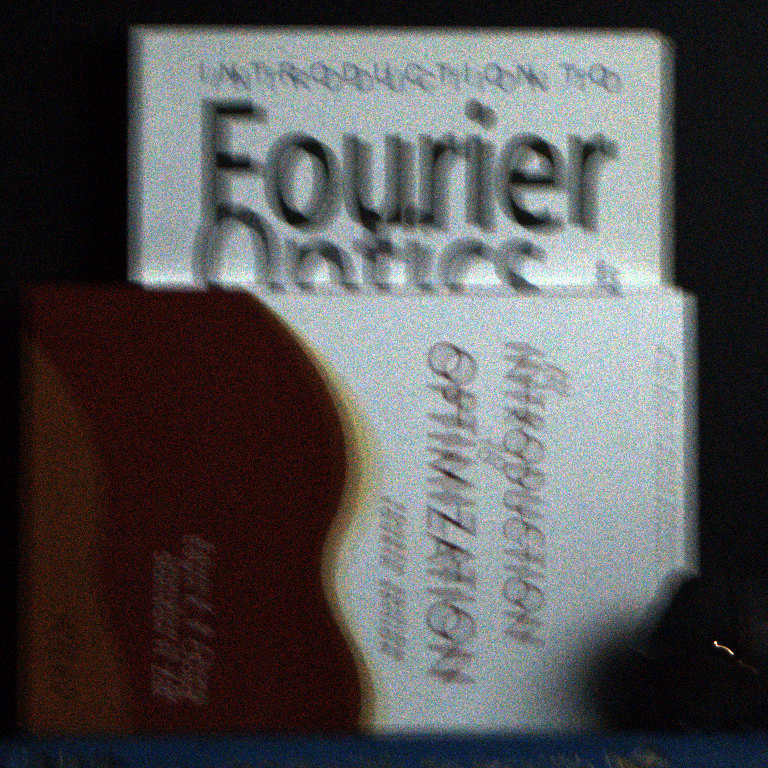} \\ Input }} &
\makecell{\includegraphics[width=0.18\linewidth]{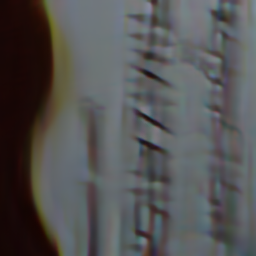} \\ SRN } &
\hspace{-1.5ex}\makecell{\includegraphics[width=0.18\linewidth]{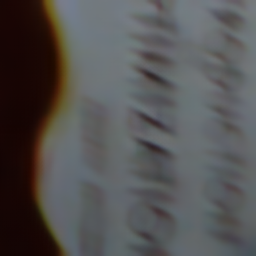} \\ DHMPN} & 
\hspace{-1.5ex}\makecell{\includegraphics[width=0.18\linewidth]{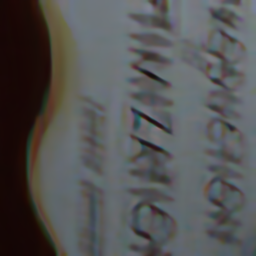} \\ MPR-Net} \\
 & \hspace{-1.5ex}\makecell{\includegraphics[width=0.18\linewidth]{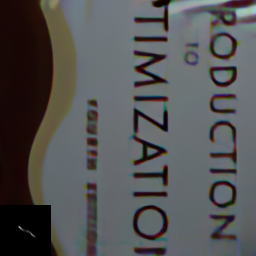} \\ \textbf{Ours}}  
 & \makecell{\includegraphics[width=0.18\linewidth]{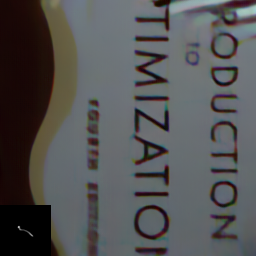} \\ Non-Blind} & \hspace{-1.5ex}\makecell{\includegraphics[width=0.18\linewidth]{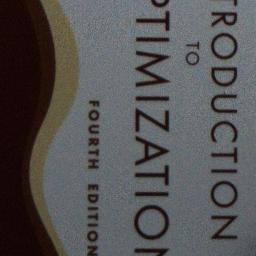} \\ Ground-Truth}   
\\
\end{tabular}
\vspace{1.0ex}
\centering
\caption{\textbf{Visual comparisons on Photon-Limited Deblurring Dataset}. Qualitative results on realistic blurred and photon-limited images from the Photon-Limited Deblurring dataset \cite{sanghvi2021photon}.The inset image for "Ours" and "Non-Blind" represent the estimated and ground-truth kernel respectively. For a more extensive set of qualitative results, we refer the reader to the supplementary document. }
\label{fig:qual_comparison_pldd}
\end{figure*}

\textbf{Color Reconstruction}
We show reconstructions on examples from the real-blur dataset in Figure \ref{fig:real_blur_examples}. While our method is grayscale, we perform colour reconstruction by estimating the kernel from the luminance-channel. Given the estimated kernel, we deblur each channel of the image using the non-blind solver and then combine the different channels into a single RGB-image. \emph{Note that all qualitative examples in this paper for end-to-end trained networks are trained using the GoPro dataset}, since they provide the better visual result.

\begin{figure}
\begin{tabular}{cccc}
    \hspace{-2.0ex}\includegraphics[width=0.24\linewidth]{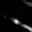}&
    \hspace{-2.0ex}\includegraphics[width=0.24\linewidth]{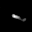}&
    \hspace{-2.0ex}\includegraphics[width=0.24\linewidth]{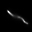}&
    \hspace{-2.0ex}\includegraphics[width=0.24\linewidth]{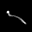}\\
    \hspace{-2.0ex}\includegraphics[width=0.24\linewidth]{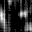}&
    \hspace{-2.0ex}\includegraphics[width=0.24\linewidth]{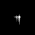}&
    \hspace{-2.0ex}\includegraphics[width=0.24\linewidth]{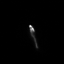}&
    \hspace{-2.0ex}\includegraphics[width=0.24\linewidth]{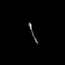}\\
    \hspace{-2.0ex}\makecell{Two-Phase \\ \cite{xu2010two}} & \hspace{-2.0ex}\makecell{Sanghvi \\ et. al  \cite{sanghvi2022photon}} & \hspace{-2.0ex}\makecell{\textbf{Ours}} & \hspace{-2.0ex}\makecell{\textbf{Ground-}\\\textbf{Truth}}  
\end{tabular}
\caption{\textbf{Estimated Kernels for different methods}: We show the estimated kernels from two examples from the PLDD dataset. Two-Phase \cite{xu2010two} uses blur-only image $G(\vy)$ as input, and ground-truth kernel is estimated using a point-source.}
\label{fig:kernels}
\end{figure}

\textbf{Photon-Limited Deblurring Dataset}
We also show qualitative examples from photon-limited deblurring dataset \cite{sanghvi2021photon} which contains  30 images' raw sensor data, blurred by camera shake and taken in extremely low-illumination. For reconstructing these images, we take the average of the R, G, B channels of the Bayer patter image, average it and then reconstruct it using the given method. The qualitative results for this dataset can be found in Figure \ref{fig:qual_comparison_pldd}. We also show the estimated kernels, along with estimated kernels from \cite{xu2010two,sanghvi2022photon}, in Figure \ref{fig:kernels}. 

However, instead of using the reblurring loss directly, we find the scheme is more numerically stable if we take the gradients of the image first and then estimate the reblurring loss. This can be explained by the fact that unlike simulated data, the photon level is not known exactly and is estimated using the sensor data itself by a simple heuristic. For further details on how to use the sensor data, we refer the reader to \cite{sanghvi2021photon}.

\subsection{Ablation Study}
In Table \ref{tab:ablation}, we provide an ablation study by running the scheme for different number of key points i.e. $K = 4, 6,$ and $8$ and without KTN ($K=0$) on RealBlur dataset. Through this study, we demonstrate the effect of the Kernel Trajectory Network has on the iterative scheme. As expected, changing the search space for kernel estimation improves the performance significantly across all metrics. Increasing the number of key points used for representing kernels also steadily improves the performance of the scheme, which can be explained by the fact there are larger degrees of freedom. 

\begin{table}[ht]
    \setlength\doublerulesep{0.5pt}
    \centering
    \scalebox{0.9}{
    \begin{tabular}{cc|cccc}
        \toprule
        \multicolumn{2}{c}{  \makecell{Key Points $\downarrow$ \\ Photon Level $\downarrow$ \\ Metric $\rightarrow$ }} & PSNR & SSIM & \makecell{LPIPS-\\Alex} & \makecell{LPIPS-\\VGG} \\
        \midrule[0.3pt]  
        \multirow{4}{*}[0.0em]{$\alpha = 10$} & \hspace{-1.0ex}$K = 0$ & 25.44 & 0.719 & 0.295 & 0.434 \\
         & \hspace{-1.0ex}$K = 4$ & 25.90 & 0.730 & 0.286 & 0.423 \\
         & \hspace{-1.0ex}$K = 6$ & 26.38 & 0.735 & 0.279 & 0.422 \\
         & \hspace{-1.0ex}$K = 8$ & \textbf{26.73} & \textbf{0.740} & \textbf{0.272} & \textbf{0.414} \\
        \midrule[0.3pt]
        \multirow{4}{*}[0.0em]{$\alpha = 20$} & \hspace{-1.0ex}$K = 0$ & 25.61 & 0.767 & 0.267 & 0.407 \\
         & \hspace{-1.0ex}$K = 4$ & 26.93 & 0.785 & 0.242 & 0.385 \\
         & \hspace{-1.0ex}$K = 6$ & 26.78 & 0.784 & 0.246 & 0.392 \\
         & \hspace{-1.0ex}$K = 8$ & \textbf{27.26} & \textbf{0.795} & \textbf{0.240} & \textbf{0.384} \\
        \midrule[0.3pt]
        \multirow{4}{*}[0.0em]{$\alpha = 40$} & \hspace{-1.0ex}$K = 0$ & 25.82 & 0.759 & 0.248 & 0.390 \\
         & \hspace{-1.0ex}$K = 4$ & 26.88 & 0.771 & 0.228 & 0.374 \\
         & \hspace{-1.0ex}$K = 6$ & \textbf{27.37} & 0.780 & 0.216 & 0.366 \\
         & \hspace{-1.0ex}$K = 8$ & 27.29 & \textbf{0.785} & \textbf{0.214} & \textbf{0.361} \\
        \midrule[0.3pt]
        \bottomrule[1pt]
    \end{tabular}
    }
    \caption{\textbf{Ablation Study for effect of Kernel Trajectory Network $T(.)$}: Reconstruction metrics for different number of key points. $K =0 $ represents the variant of the scheme which does not use the Kernel-Trajectory Network and estimates the kernel directly from Stage II.}
    \label{tab:ablation}
\end{table}

\section{Conclusion}
In this paper, we use an iterative framework for the photon-limited blind deconvolution problem. More specifically, we use a non-blind solver which can deconvolve Poisson corrupted and blurred images given a blur kernel. To mitigate ill-posedness of the kernel estimation in such high noise regimes, we propose a novel low-dimensional representation to represent the motion blur. By using this novel low-dimensional and differentiable representation as a search space, we show state-of-the-art deconvolution performance and outperform end-to-end trained image restoration networks by a significant margin.

We believe this is a promising direction of research for both deblurring and general \emph{blind inverse problems} i.e., inverse problems where the forward operator is not fully known. Future work could involve a supervised version of this scheme which does not involve backpropagation through a network as it would greatly reduce the computational cost. Another possible line of research could be to apply this framework to the problem of spatially varying blur. 

\subsection*{Acknowledgement}
The work is supported, in part, by the United States National Science Foundation under the grants IIS-2133032 and ECCS-2030570.

{\small
\bibliographystyle{ieee_fullname}
\bibliography{ref}
}

\end{document}

\title{Structured Kernel Estimation for Photon-Limited Deconvolution - Supplementary}

\author{Yash Sanghvi, Zhiyuan Mao, Stanley H. Chan\\
School of Electrical and Computer Engineering, Purdue University\\
{\tt\small \{ysanghvi, mao114, stanchan\}@purdue.edu}}
\maketitle
\section{Overview of Camera Noise Model}
In this section, we present an overview of different camera noise sources followed by a justification of the Poisson noise model used for photon-limited settings. Consider the sensor output of $i^{\text{th}}$ pixel of camera, denoted as $Y_i$. From \cite{zhang2021rethinking}, we model $Y_i$ as the following random variable:
\begin{align} 
Y_i \sim K_d \Big( K_a (\mathcal{P}(I_i) + \eta_{a}) + \eta_d + \eta_q\Big) \label{eq:noise_model}
\end{align}
Here $I_i$ represents the average number of incident photons during the exposure. $K_a$, $K_d$ represent the analog and digital gain respectively. $\eta_a$ represents noise sources before the analog gain (dark current shot noise, flicker noise etc.) and $\eta_d$ represents noise sources before the digital gain (thermal noise, fixed pattern noise). $\eta_q$ represent the quantization noise. From Eq. \eqref{eq:noise_model}, we can see view $Y_i$ is a noisy measurement of the parameter $I_i$. 

Eq. \eqref{eq:noise_model} can be simplified in the following way:
\begin{align}
    \tilde{Y_i} \sim \underset{\text{signal-dependent}}{\underbrace{\mathcal{P}(I_i)}} + \underset{\text{signal-independent}}{\underbrace{\eta_a + \frac{1}{K_a}(\eta_d + \eta_q)}}
\end{align}
where $\tilde{Y_i} \bydef Y_i/(K_a K_d)$ is the normalized the pixel measurement. The noise sources can be easily decoupled into signal dependent Poisson noise and signal independent noise sources. The latter is often approximated as zero mean Gaussian noise in literature \cite{foi2008practical, foi2009clipped, luisier2010image} leading to the following Poisson-Gaussian mixture modelling
\begin{align}
    \tilde{Y_i} \sim \mathcal{P}(I_i) + Z_i \;\;\;\;\;\;\;\; Z_i \sim \mathcal{N}(0, \sigma^2)
\end{align}
We can further break down average number of incident photon $I_i $ as $I_i \bydef \alpha x_i$ where $\alpha$ is a function of camera parameters such as exposure time, quantum efficiency, sensor area etc. and $x_i$ is the scene radiance corresponding to the $i^{\text{th}}$ pixel. This helps us decouple the scene and camera characteristics in the signal $I_i$. Therefore, we arrive at the following camera model formulation where $Y_i$ is the measurement of the true signal $x_i$:
\begin{align}
    \tilde{Y_i} \sim \mathcal{P}(\alpha x_i) + \mathcal{N}(0, \sigma^2) \label{eq:poisson_and_gauss}
\end{align}
 
\subsection{Poisson Noise SNR}
For this subsection, we ignore the Gaussian term in Eq. (\ref{eq:poisson_and_gauss}) i.e. $\sigma =0$ to focus on the nature of Poisson noise. An interesting property of the Poisson random variable is that its mean and variance are the same. Thus, the signal-to-noise ratio for $\tilde{Y_i}$, in absence of Gaussian noise, is given as follows:
\begin{align}
    \text{SNR}(\tilde{Y_i}) \bydef \frac{\mathbb{E}[\tilde{Y_i}]}{\sqrt{\Var[\tilde{Y_i}]}} =\sqrt{\alpha x_i}
\end{align}
This implies that our measurements get noisier with decreasing number of incident photons. If the scene is not well-illuminated (low $x_i$) or the exposure is short (small $\alpha$), the number of incident photons on the sensor is also low leading to noisier images.    

\subsection{Photon-Limited Scenes}
Scenes where the Poisson noise dominates other sources of noise in the measurements are defined as \emph{photon-limited} \cite{hasinoff2014photon}. For the random variable $\tilde{Y_i}$ defined in \eqref{eq:poisson_and_gauss}, this occurs when the variance due to Poisson noise is greater than the variance due to Gaussian noise, i.e. $\alpha x_i \geq \sigma^2$. 

However, for the purpose of this paper, not all photon-limited scenes are equally significant. To emphasize this point, we inspect the SNR for Poisson-Gaussian mixture $\tilde{Y_i}$ for different levels of $\alpha$. The SNR for $\tilde{Y_i}$ in presence of both Poisson and Gaussian noise is given as follows:
\begin{align}
    \text{SNR}[\tilde{Y_i}] \bydef \frac{\mathbb{E}[\tilde{Y_i}]}{\sqrt{\Var[\tilde{Y_i}]}} = \frac{\alpha x_i}{\sqrt{\alpha x_i + \sigma^2}}
\end{align}

Consider the case of read noise $\sigma = 1.6$e- \cite{ma2022ultra}. We assume $x_i=1$ and inspect the random variable $Y_i$ for different $\alpha$ in the photon-limited regime, as shown in Table \ref{tab:diff_alpha}.

\begin{table}[h]
    \centering
    \begin{tabular}{|c|c|}
        \toprule
        Photon level $\downarrow$  & SNR (in dB)  \\
        \hline
        $\alpha = 1000$ & 29.98 dB \\
        $\alpha = 40$ & 15.75 dB \\
        $\alpha = 20$ & 12.48 dB \\
        $\alpha = \sigma^2$ & 1.07 dB \\
        \bottomrule
    \end{tabular}
    \caption{Signal-to-noise ratio (SNR) for different photon levels in photon-limited regime.}
    \label{tab:diff_alpha}
\end{table}

From the table we can conclude the following: Images taken in well-illuminated scenes with good exposure can be approximated as noiseless for the purpose of deblurring. However, on the other end of photon-limited regime i.e. $\alpha=\sigma^2$, there is too much noise in the image for any meaningful recovery from a single frame. Therefore, for this paper, we explore photon-limited deconvolution for $\alpha \in [10,40]$ where shot noise dominates read noise but there is still a possibility of extracting the clean image.

\section{Initialization Algorithm}
In Algorithm \ref{alg:initialize}, we describe the kernel initialization method for the proposed method. This scheme is a minor variation of the kernel estimation method from \cite{delbracio2021polyblur} and used to estimate a rectilinear kernel with parameters $\{\rho, \theta\}$ from the blur-only image $G(\vy)$. We would like to reiterate to the reader that the scheme in Algorithm \ref{alg:initialize} is not the kernel estimation process in its entirety and only represents the initialization process. The kernel estimated at the end of this algorithm is further refined in Stage I and II of the main iterative scheme. 
\begin{algorithm}[h]
\begin{algorithmic}[1]
\State \textbf{Input}: Blur-only Image $G(\vy)$, Number of Key Points $K$
\State Estimate gradient images $\mD_{x}, \mD_{y}$ from $G(\vy)$
\For{$\theta = 1, 2, \cdot\cdot\cdot, 180$}        \State $\mD_{\theta} \leftarrow \mD_{x}\cos(\theta) - \mD_{x}\sin(\theta) $
    \State $f_{\theta} \leftarrow \max (|\mD_{\theta}|)$
    \EndFor
\State $f_{\hat{\theta}}, \hat{\theta} \leftarrow \min f_{\theta} $
\State $\rho \leftarrow C_1\sqrt{\frac{C_0^2}{f_{\hat{\theta}}^2} - \sigma_b^2}$
\State $(x_0, y_0) \leftarrow (0,0)$
\For{$k = 1, 2, \cdot\cdot\cdot, K-1$}         \State $(x_k, y_k) \leftarrow \frac{k\rho}{K-1}\cos(\hat{\theta}), \frac{k\rho}{K-1}\sin(\hat{\theta})$
\EndFor
\State $\vz_0 \leftarrow [x_0,\; y_0,\; x_1,\; y_1,\; ...,\; x_{K-1},\; y_{K-1}]^T$ 
\State return $\vz_0$
\end{algorithmic}
\caption{Initialization for Kernel Estimation }
\label{alg:initialize}
\end{algorithm}

\section{Qualitative Comparison}
\begin{figure*}[h]
    \begin{tabular}{cccc}
    \hspace{-2.0ex}\multirow{2}[2]{*}[17mm]{\makecell{\includegraphics[width=0.39\linewidth]{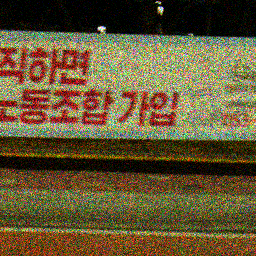} \\ Input} } & \makecell{\includegraphics[width=0.18\linewidth]{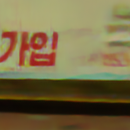} \\ SRN }&
    \makecell{\includegraphics[width=0.18\linewidth]{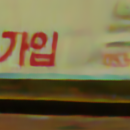} \\ DHMPN}&
    \makecell{\includegraphics[width=0.18\linewidth]{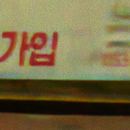} \\ MIMO-UNet+}
    \\
    & \makecell{\includegraphics[width=0.18\linewidth]{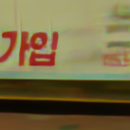} \\ MPR-Net} &
    \makecell{\includegraphics[width=0.18\linewidth]{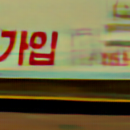} \\ \textbf{Ours}} &
    \makecell{\includegraphics[width=0.18\linewidth]{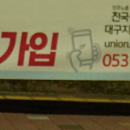} \\ Ground Truth}   
    \end{tabular}

    \begin{tabular}{cccc}
    \hspace{-2.0ex}\multirow{2}[2]{*}[17mm]{\makecell{\includegraphics[width=0.39\linewidth]{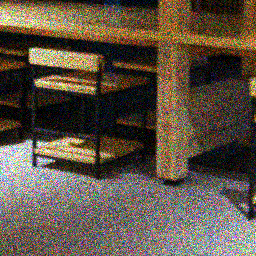} \\ Input} } & \makecell{\includegraphics[width=0.18\linewidth]{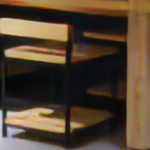} \\ SRN }&
    \makecell{\includegraphics[width=0.18\linewidth]{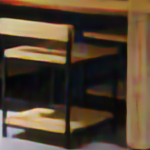} \\ DHMPN}&
    \makecell{\includegraphics[width=0.18\linewidth]{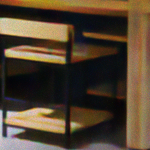} \\ MIMO-UNet+}
    \\
    & \makecell{\includegraphics[width=0.18\linewidth]{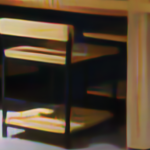} \\ MPR-Net} &
    \makecell{\includegraphics[width=0.18\linewidth]{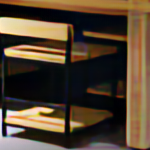} \\ \textbf{Ours}} &
    \makecell{\includegraphics[width=0.18\linewidth]{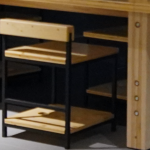} \\ Ground Truth}   
    \end{tabular}
    \caption{\textbf{ More Qualitative examples on the Real-Blur Dataset}}
\label{fig:real_blur_example1}
\end{figure*}

\begin{figure*}[h]
\centering
 \begin{tabular}{cccc}
 \hspace{-2.0ex}\multirow{2}[2]{*}[17mm]{\makecell{\includegraphics[width=0.39\linewidth]{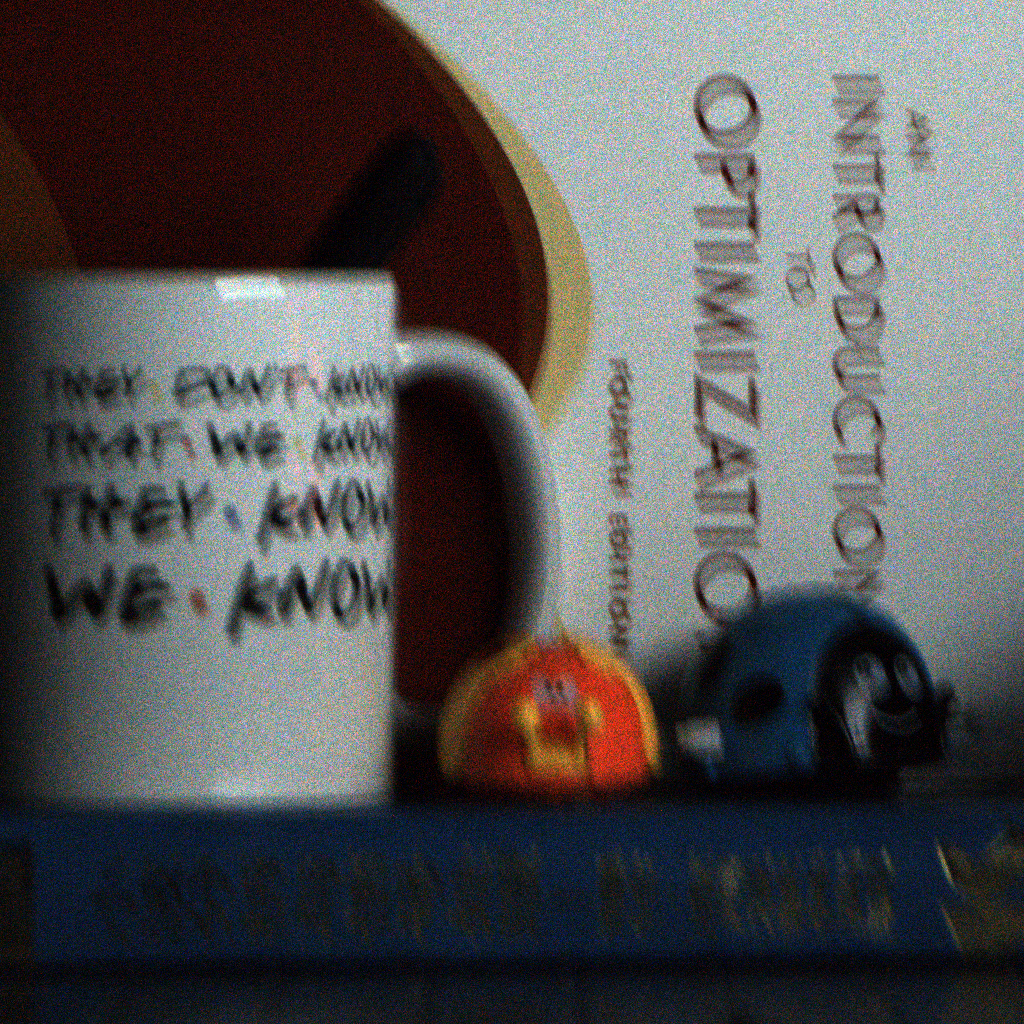} \\ Input }} &
\makecell{\includegraphics[width=0.18\linewidth]{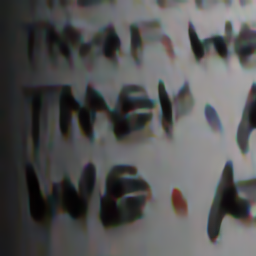} \\ SRN } &
\hspace{-1.5ex}\makecell{\includegraphics[width=0.18\linewidth]{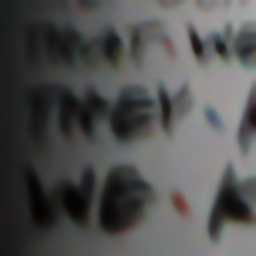} \\ DHMPN} & 
\hspace{-1.5ex}\makecell{\includegraphics[width=0.18\linewidth]{Figures/supp/figure5/srn6.png} \\ MPR} \\
 & \hspace{-1.5ex}\makecell{\includegraphics[width=0.18\linewidth]{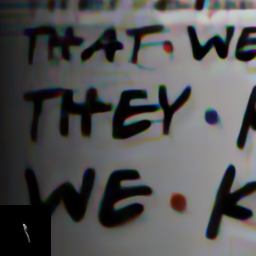} \\ \textbf{Ours}}  
 & \makecell{\includegraphics[width=0.18\linewidth]{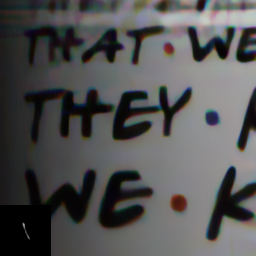} \\ Non-Blind} & \hspace{-1.5ex}\makecell{\includegraphics[width=0.18\linewidth]{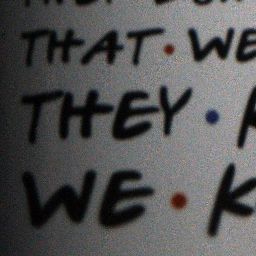} \\ Ground-Truth}   
\\
\end{tabular}
\begin{tabular}{cccc}
 \hspace{-2.0ex}\multirow{2}[2]{*}[17mm]{\makecell{\includegraphics[width=0.39\linewidth]{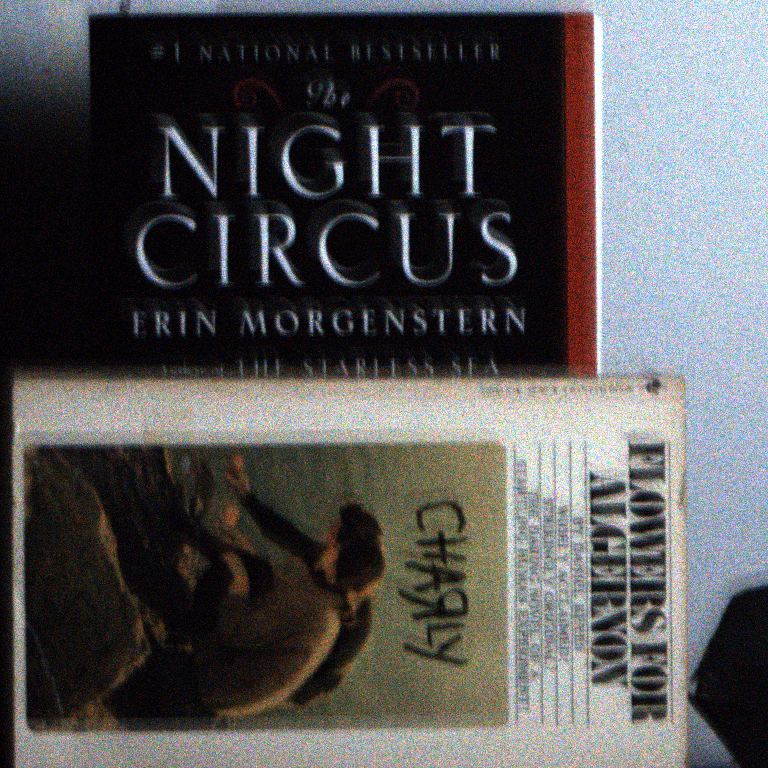} \\ Input }} &
\makecell{\includegraphics[width=0.18\linewidth]{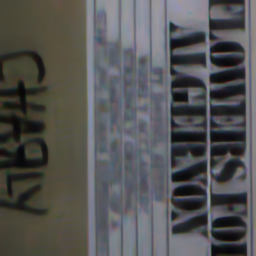} \\ SRN } &
\hspace{-1.5ex}\makecell{\includegraphics[width=0.18\linewidth]{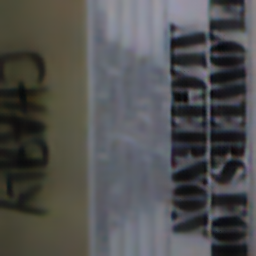} \\ DHMPN} & 
\hspace{-1.5ex}\makecell{\includegraphics[width=0.18\linewidth]{Figures/supp/figure6/srn2.png} \\ MPR} \\
 & \hspace{-1.5ex}\makecell{\includegraphics[width=0.18\linewidth]{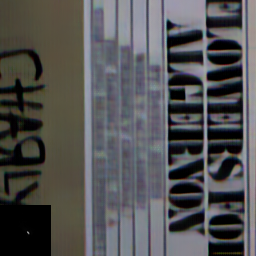} \\ \textbf{Ours}}  
 & \makecell{\includegraphics[width=0.18\linewidth]{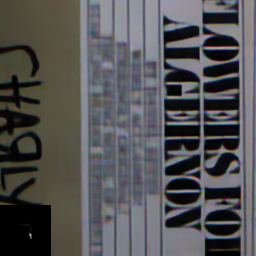} \\ Non-Blind} & \hspace{-1.5ex}\makecell{\includegraphics[width=0.18\linewidth]{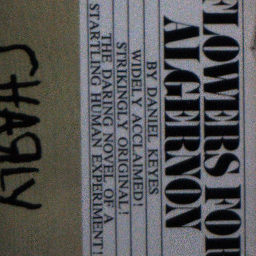} \\ Ground-Truth}   
\\
\end{tabular}
\caption{\textbf{More comparisons on Photon-Limited Deblurring Dataset} \cite{sanghvi2021photon}.}
\label{fig:qual_comparison_pldd}
\end{figure*}

\begin{figure*}[h]
\centering
\begin{tabular}{cccc}
 \hspace{-2.0ex}\multirow{2}[2]{*}[17mm]{\makecell{\includegraphics[width=0.39\linewidth]{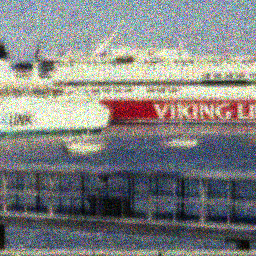} \\ Input }} &
\makecell{\includegraphics[width=0.18\linewidth]{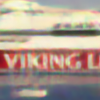} \\ SRN } &
\hspace{-1.5ex}\makecell{\includegraphics[width=0.18\linewidth]{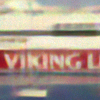} \\ MIMO-UNet++} & 
\hspace{-1.5ex}\makecell{\includegraphics[width=0.18\linewidth]{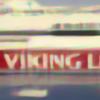} \\ MPR-Net} \\
 & \hspace{-1.5ex}\makecell{\includegraphics[width=0.18\linewidth]{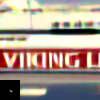} \\ \textbf{Ours}}  
 & \makecell{\includegraphics[width=0.18\linewidth]{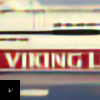} \\ Non-Blind} & \hspace{-1.5ex}\makecell{\includegraphics[width=0.18\linewidth]{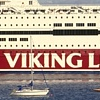} \\ \textbf{Ground-Truth}}   
\\
\end{tabular}
\begin{tabular}{cccc}
 \hspace{-2.0ex}\multirow{2}[2]{*}[17mm]{\makecell{\includegraphics[width=0.39\linewidth]{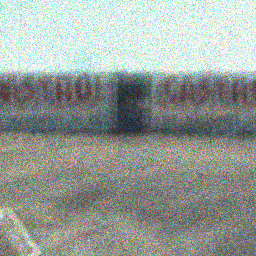} \\ Input }} &
\makecell{\includegraphics[width=0.18\linewidth]{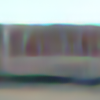} \\ SRN } &
\hspace{-1.5ex}\makecell{\includegraphics[width=0.18\linewidth]{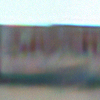} \\ MIMO-UNet++} & 
\hspace{-1.5ex}\makecell{\includegraphics[width=0.18\linewidth]{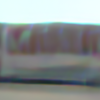} \\ MPR-Net} \\
 & \hspace{-1.5ex}\makecell{\includegraphics[width=0.18\linewidth]{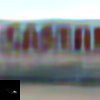} \\ \textbf{Ours}}  
 & \makecell{\includegraphics[width=0.18\linewidth]{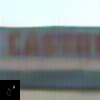} \\ Non-Blind} & \hspace{-1.5ex}\makecell{\includegraphics[width=0.18\linewidth]{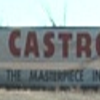} \\ \textbf{Ground-Truth}}   
\end{tabular}
\caption{\textbf{Qualitative Examples on Synthetic Blur}: "Non-Blind" is provided for reference and serves as an upper bound on the deconvolution performance. It is obtained through \textit{PhD-Net} with noisy-blurred image and ground truth kernel as inputs. The kernels in inset of "Ours" and "Non-Blind" represent the estimated and true blur kernel respectively. }
\label{fig:synthetic_examples}
\end{figure*}

Extended qualitative results comparing end-to-end trained methods to our approach are provided in the supplementary document. Figure \ref{fig:real_blur_examples} and \ref{fig:real_blur_example1} provide qualitative examples from the RealBlur dataset which contains realistic blur. Figure \ref{fig:qual_comparison_pldd} contains examples from the PLDD dataset \cite{sanghvi2022photon} cotnains real-shot noise corrupted and blurred image sensor data along with the ground-truth kernel, as measured using a point source. Finally, Figure \ref{fig:synthetic_examples} provides reconstruction results on synthetically blurred images using motion kernels from Levin dataset \cite{levin2006blind}.
\section{KTN Architecture and Training}
\begin{figure*}
    \centering
    \includegraphics[width=0.95\linewidth]{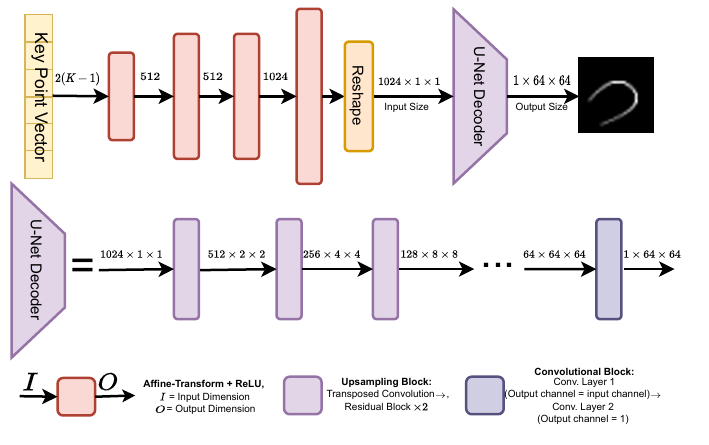}
    \caption{\textbf{Kernel Trajectory Network Architecture} 3 fully connected layers followed by a U-Net decoder}
    \label{fig:arch_ktn}
\end{figure*}

The KTN architecture can be summarized as follows: the vectorized control points of dimension $2\times(K-1)$ are passed through 3 fully-connected layers followed by reshaping into an image. The reshaped image is then passed through the decoding half of a UNet to give the kernel output. The final output, when used in the iterative scheme, is clipped to zero and then normalized to one. Architecture details of KTN are provide in Figure \ref{fig:arch_ktn}.
\section{Implementation Details}
\textbf{Boundary Conditions}: While blurring the image synthetically, the boundary conditions are important to take into account. Circular boundary conditions allow the blur operator to be written in terms of FFTs and making it computationally inexpensive, it is not a realistic assumption for natural blur. A more appropriate boundary condition to assume is \emph{symmetric boundary condition}. 

This has major implications for our inverse problem scheme. Since PhD-Net assumes circular boundary conditions, we need to pad the image symmetrically, pass through PhD-Net and crop out the relevant portion to deblur the image without any artifacts. Second, when calculating the reblurring loss, $\vh_{\vz} \circledast F(\vy, \vh_{\vz})$ needs to be calculated using symmetric boundary conditions. 

\textbf{Step Size and Backtracking}: For Stage I, we set the initial step size as $\delta = 10^5$ and for Stage II, we set $\delta = 2.0$. For every iteration, we check whether the current choice of step-size decreases the cost-function or not. If it doesn't, then the step-size is reduced by half for rest of the iterative scheme until the next time the cost function increases instead of decreasing. Note that $\delta$ is set very large in the first stage compared to the second.  This is because the gradients are backpropagated through two networks i.e., $F(.)$ and $T(.)$ instead of one, leading to the vanishing gradient problem and hence justifying the larger step size.

\textbf{Computational Time}: For each stage of the iterative scheme, we limit the number of iterations to 150. The experiments in the main document are performed on a Nvidia TitanX GPU, and take approximately 0.35 seconds per iteration.

\begin{figure*}[h]
    \centering
    \begin{tabular}{cccc}
    \hspace{-2.0ex}\multirow{2}[2]{*}[17mm]{\makecell{\includegraphics[width=0.39\linewidth]{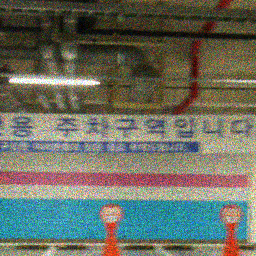} \\ Input} } & \makecell{\includegraphics[width=0.18\linewidth]{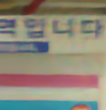} \\ SRN }&
    \makecell{\includegraphics[width=0.18\linewidth]{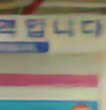} \\ DHMPN}&
    \makecell{\includegraphics[width=0.18\linewidth]{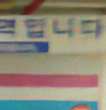} \\ MIMO-UNet+}
    \\
    & \makecell{\includegraphics[width=0.18\linewidth]{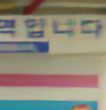} \\ MPR-Net} &
    \makecell{\includegraphics[width=0.18\linewidth]{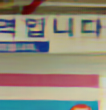} \\ \textbf{Ours}} &
    \makecell{\includegraphics[width=0.18\linewidth]{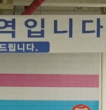} \\ Ground Truth}          
    \end{tabular}

    \begin{tabular}{cccc}
    \hspace{-2.0ex}\multirow{2}[2]{*}[17mm]{\makecell{\includegraphics[width=0.39\linewidth]{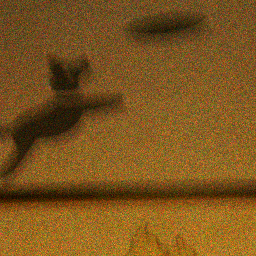} \\ Input} } & \makecell{\includegraphics[width=0.18\linewidth]{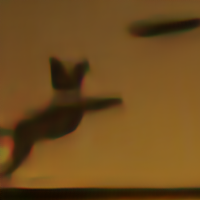} \\ SRN }&
    \makecell{\includegraphics[width=0.18\linewidth]{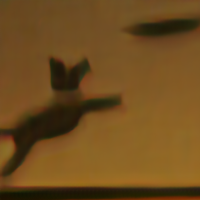} \\ DHMPN}&
    \makecell{\includegraphics[width=0.18\linewidth]{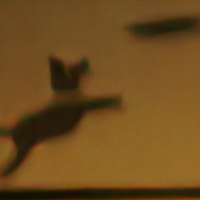} \\ MIMO-UNet+}
    \\
    & \makecell{\includegraphics[width=0.18\linewidth]{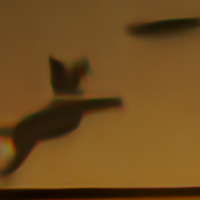} \\ MPR-Net} &
    \makecell{\includegraphics[width=0.18\linewidth]{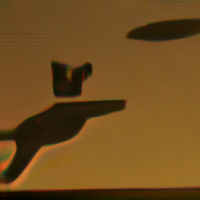} \\ \textbf{Ours}} &
    \makecell{\includegraphics[width=0.18\linewidth]{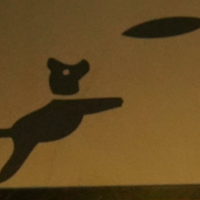} \\ Ground Truth}          
    \end{tabular}
    \caption{\textbf{Qualitative examples on the Real-Blur Dataset}}
    \label{fig:real_blur_examples}
\end{figure*}

{\small
\bibliographystyle{ieee_fullname}
\bibliography{ref}
}

%% file: commands.tex
\usepackage{setspace}
\usepackage{graphicx}
\usepackage{multirow,multicol}
\usepackage[]{amsmath}
\usepackage{amsbsy}
\usepackage{amssymb}
\usepackage{amsfonts}
\usepackage{pifont}
\usepackage{balance}
\usepackage{fancyvrb}

{\begin{list}               %    with flush left bullets.
    {$\bullet$ \hfill}{
        \setlength{\leftmargin}{\parindent}
        \setlength{\parsep}{0.04\baselineskip}
        \setlength{\itemsep}{0.5\parsep}
        \setlength{\labelwidth}{\leftmargin}
        \setlength{\labelsep}{0em}}
    }
{\end{list}}

\providecommand{\cref}[1]{Chapter~\ref{#1}}

\providecommand{\R}{\ensuremath{\mathbb{R}}}

\providecommand{\bydef}{\overset{\text{def}}{=}}

\providecommand{\calL}{\mathcal{L}}

\providecommand{\calS}{\mathcal{S}}

\providecommand{\mI}{\mathbf{I}}

\providecommand{\vh}{\mathbf{h}}

\providecommand{\vv}{\mathbf{v}}

\providecommand{\vx}{\mathbf{x}}
\providecommand{\vy}{\mathbf{y}}
\providecommand{\vz}{\mathbf{z}}

\newcommand{\cmark}{\ding{51}}
\newcommand{\xmark}{\ding{53}}

\newcommand\Tstrut{\rule{0pt}{2.6ex}}         % = `top' strut
  